\documentstyle[12pt]{article}
\def\be {\begin{equation}}
\def\ee {\end{equation}}
\def\ba {\begin{eqnarray}}
\def\ea {\end{eqnarray}}

\def\bi {\begin{itemize}}
\def\ei {\end{itemize}}
\begin{document}
\def\bea{\begin{eqnarray}}
\def\eea{\end{eqnarray}}
\title{\bf {Semiclassical Corrections to the Cardy-Verlinde Formula of Kerr Black Holes }}
 \author{M.R. Setare  \footnote{E-mail: rezakord@ipm.ir}
  \\{Department of Science,  Payame Noor University. Bijar. Iran}}
\date{\small{}}

\maketitle
\begin{abstract}
In this letter, we compute the corrections to the Cardy-Verlinde
formula of $4-$dimensional Kerr black hole. These corrections are
considered within the context of KKW analysis and arise as a result
of the self-gravitational effect. Then we show, one can taking into
account the semiclassical corrections of the Cardy-Verlinde entropy
formula by just redefining the Virasoro operator $L_0$ and the
central charge $c$.
 \end{abstract}

\newpage

\section{Introduction}
Concerning the quantum process called Hawking effect \cite{hawking1}
much work has been done using a fixed background during the emission
process. The idea of Keski-Vakkuri, Kraus and Wilczek (KKW)
\cite{KKW1}-\cite{KKW5} is to view the black hole background as
dynamical by treating the Hawking radiation as a tunnelling process.
The energy conservation is the key to this description. The total
(ADM) mass is kept fixed while the mass of the black hole under
consideration decreases due to the emitted radiation. The effect of
this modification gives rise to additional terms in the formulae
concerning the known results for black holes
\cite{corrections1}-\cite{corrections6}; a non-thermal partner to
the thermal spectrum of the Hawking radiation
shows up.\\
The Cardy-Verlinde formula proposed
   by Verlinde \cite{Verl}, relates the entropy of a  certain CFT with its total
energy and its Casimir energy in arbitrary dimensions. Using the
AdS$_{d}$/CFT$_{d-1}$ \cite{AdS1}-\cite{AdS4} and
dS$_{d}$/CFT$_{d-1}$ correspondences \cite{{AS1},{AS2},{AS3}} , this
formula has been shown to hold exactly for different black holes
(see for example {\cite{odi1}-\cite{set2}}). All cases where the
Cardy-Verlinde formula has been shown to hold up in these references
had as a necessary ingredient a negative cosmological constant, or,
more generally, a certain potential term for supergravity scalars.
This guarantees that the theory admits AdS vacua, and thus one has a
dual description in terms of a conformal field theory on the
boundary of AdS. A natural question is whether the Cardy-Verlinde
formula holds in a more general setting, e.~g.~for black holes that
are asymptotically flat rather than approaching AdS space. The
authors of \cite{klem} have  shown
that this is indeed the case.  \\
In this paper, in the context of KKW analysis we consider the Kerr
black hole as a dynamical background and  we derive the modified
event horizon radius of Kerr black hole. Then we derive the
corresponding modified thermodynamic quantities of the Kerr black
hole and in the lowest order of the emitted shell of energy. Then we
consider the generalized Cardy-Verlinde formula of a $4-$dimensional
Kerr black hole. We obtain the semiclassical corrections to this
entropy formula. Finally we obtain the corrections to the quantities
entering the Cardy-Verlinde formula:Virasoro operator and the
central charge.
\section{Thermodynamical quantities of Kerr black hole}
We consider the asymptotically flat Kerr black holes
\cite{Myers:1986un}
\begin{eqnarray}
ds^2 &=& -\frac{\Delta}{\rho^2}[dt - a\sin^2\theta d\phi]^2
         + \frac{\rho^2}{\Delta}dr^2 + \rho^2 d\theta^2 \nonumber \\
     & & + \frac{\sin^2\theta}{\rho^2}[adt - (r^2+a^2)d\phi]^2
        , \label{kerr}
\end{eqnarray}
where
\begin{eqnarray}\label{delta}
\Delta &=& (r^2 + a^2) - 2Mr, \nonumber \\
\rho^2 &=& r^2 + a^2\cos^2\theta.
\end{eqnarray}
The event horizon $r=r_+$ is given by \be \label{hor}
r_+=M+\sqrt{M^{2}-a^2} \ee

 The temperature, entropy, energy, angular momentum and angular velocity of
the horizon are as following\cite{Myers:1986un} \be \label{tem}
T=\frac{r_{+}^{2}-a^2}{4\pi r_+ (r_{+}^{2}+a^2)} \ee \be
\label{ent}S_{BH} = \frac{V_2}{4G}(r_+^2+a^2), \ee \be \label{en}E =
\frac{V_2}{8\pi G r_+}(r_+^2+a^2) \ee \be \label{ang}J =
\frac{aV_2}{8\pi G r_+}(r_+^2+a^2) \ee \be \label{angv}\Omega_{H} =
\frac{a}{r_+^2+a^2} \ee where $V_2$ denotes the volume of the unit
$2$-sphere.

\section{Self-Gravitational Corrections to Cardy-Verlind Formula}
The entropy of a $(1+1)-$dimensional CFT is given by the well-known
Cardy formula \cite{cardy} \be
S=2\pi\sqrt{\frac{c}{6}(L_0-\frac{c}{24})}, \label{car} \ee where
$L_0$ represent the product $ER$ of the energy and radius, and the
shift of $\frac{c}{24}$ is caused by the Casimir effect. After
making the appropriate identifications for $L_0$ and $c$, the same
Cardy formula is also valid for CFT in arbitrary spacetime
dimensions $d$ in the form \cite{Verl} \be S_{CFT}=\frac{2\pi
R}{d-2}\sqrt{E_c(2E-E_c)}, \label{Cardy}
 \ee the so called Cardy-Verlinde formula, where $R$ is the radius of the
 system, $E$ is the total energy and $E_{C}$ is the Casimir energy.
  The definition of the Casimir energy is derived by the
violation of the Euler relation as \cite{Verl}
\begin{equation}\label{cas1}
E_C = 2(E - TS_{BH} + pV - \Omega_{H} J),
\end{equation}
 the pressure of the CFT is defined as $p=E/2V$, where $V$ is the volume of the system. The total
energy may be written as the sum of two terms \be E(S, V)=E_{E}(S,
V)+\frac{1}{2}E_{C}(S, V)\label{ext}\ee where $E_{E}$ is the purely
extensive part of the total energy $E$ and the Casimir energy, is
the subextensive part of $E$. Using the equation of state, $pV =
E/2$, and substituting Eqs.(\ref{tem})-(\ref{angv}) into
Eq.(\ref{cas1}), we get
\begin{equation}\label{cas2}
E_C = \frac{V_2}{4\pi G r_+}(r_+^2+a^2) = 2E.
\end{equation}
So far, mostly asymptotically AdS and dS
 black hole solutions have been considered \cite{AdS1}-\cite{set2}. In \cite{klem},
 it is shown that even the Schwarzschild and Kerr black hole solutions, which are
 asymptotically flat, satisfy the modification of the Cardy-Verlinde formula
 \be S_{CFT}=\frac{2\pi R}{d-2}\sqrt{2EE_c}. \label{cardy1} \ee This result holds also
 for various charged black hole solution with asymptotically flat spacetime
 \cite{yum}.\\
 We are interested primarily in the corrections to the entropy (\ref{ent}) that arise in the context of KKW
analysis \cite{KKW1}-\cite{KKW5} . Let us remind that the key point
to the KKW analysis is that the total energy of the spacetime under
study is kept fixed while the black hole mass is allowed to vary. To
calculate the emission rate, we should introduce the
Painlev$\acute{\mathrm{e}}$-Ker coordinate system. The line element
in the Painlev$\acute{\mathrm{e}}$-Kerr coordinate system is given
in Ref. \cite{Zhang2}. Namely, \bea
ds^{2}  &  =\widehat{g}_{00}dt^{2}+2\sqrt{\widehat{g}_{00}(1-g_{11}%
)}dtdr+dr^{2}+[\widehat{g}_{00}G(r,\theta)^{2}+g_{22}]d\theta^{2}\nonumber
\\
&  +2\widehat{g}_{00}G(r,\theta)dtd\theta+2\sqrt{\widehat{g}_{00}(1-g_{11}%
)}G(r,\theta)drd\theta,\label{Painleve Kerr-Newman} \eea where
\begin{equation}
\widehat{g}_{00}=-\frac{\rho^{2}\Delta}{(r^{2}+a^{2})^{2}-\Delta
a^{2}\sin ^{2}\theta},
\end{equation}%
\begin{equation}
g_{11}=\frac{\rho^{2}}{\Delta}, \hspace{1cm}g_{22}=\rho^{2}.
\end{equation}%
and $G(r,\theta)$ is given by
\begin{equation}
G(r,\theta)=\int\frac{\partial
F(r,\theta)}{\partial\theta}dr+C(\theta),
\end{equation}
also $F(r,\theta)$ satisfies
\begin{equation}
g_{11}+\widehat{g}_{00}F(r,\theta)^{2}=1, \label{demend}%
\end{equation}
here, $C(\theta)$ is an arbitrary analytic function of $\theta$. The
outgoing radial null geodesics followed by the massless particles,
i.e. the shell of energy, is given by
\begin{equation}
\dot{r}
=\frac{\Delta}{2}\sqrt{\frac{\rho^2}{(\rho^2-\Delta)[(r^{2}+a^{2})^{2}-\Delta
a^{2}\sin ^{2}\theta ]}}. \label{vp}
\end{equation}
To calculate the emission rate correctly, we should take into
account the self-gravitation of the tunnelling particle with energy
$\omega$. That is, we should replace $M$ with $M-\omega$ in
(\ref{Painleve Kerr-Newman}). The coordinate $\varphi$ does not
appear in the line element expressions (\ref{Painleve Kerr-Newman}).
That is to say, $\varphi$ is an ignorable coordinate in the
Lagrangian function $L$. To eliminate this freedom completely, the
action for the classically forbidden trajectory should be written as
\begin{equation}
S=\int\nolimits_{t_{i}}^{t_{f}}(L-P_{\varphi}\dot{\varphi})dt,
\end{equation}
which is related to the emission rate of the tunnelling particle by
\begin{equation}
\Gamma\sim e^{-2Im S}.
\end{equation}
Therefore, the imaginary part of the action is
\begin{equation}
\label{ims1}
 Im S=Im(\int_{r_{i}}^{r_{f}}
[P_{r}-\frac{P_{\varphi}\dot{\varphi}}{\dot{r}} ]dr)=Im
(\int_{r_{i}}^{r_{f}}[\int_{(0,0)}
^{(P_{r},P_{\varphi})}dP_{r}^{^{\prime}}-\frac{\dot{\varphi}}{\dot{r}}dP_{\varphi}^{^{\prime}}]dr),
\end{equation}
 where  $P_{\varphi}$ is the canonical momentum
conjugate to $\varphi$. If we treat the black hole as a rotating
sphere and consider the particle self-gravitation, we have
\begin{equation}
\dot{\varphi}=\Omega'_H,
\end{equation}
and
\begin{equation}
J'=(M-\omega ')a=P'_{\varphi},
\end{equation}
where $\Omega'_H$ is the dragged angular velocity of the event
horizon. The imaginary part of the action can be rewritten as
\begin{equation}
Im S=Im\{\int\nolimits_{r_{i}}^{r_{f}}[\int_{(0,J)}
^{(P_{r},J-\omega
a)}dP_{r}^{^{\prime}}-\frac{\Omega'_H}{\dot{r}}dJ']dr\}.
\label{ims2}
\end{equation}
We now eliminate the momentum in favor of energy by using Hamilton's
equation
\begin{equation}
\dot{r}=\frac{dH}{dP_{r}}\mid_{(r;\varphi,P_{\varphi})}=\frac{d(M-\omega')}{dP_{r}}=\frac{dM'}{dP_{r}}, \label{r2}%
\end{equation}%
Based on similar discussion to \cite{KKW1}-\cite{KKW5} , it follows
directly that a particle tunnelling across the event horizon sees
the effective metric of Eq. (\ref{Painleve Kerr-Newman}), although
with the replacements $M\rightarrow M-\omega'$. The same
substitutions in Eq. (\ref{vp}) yield the desired expression of
$\dot{r}$ as a function of $\omega'$. Thus, we can rewrite
(\ref{ims2}) in the following explicit manner
\begin{equation}
Im S=Im\int\nolimits_{r_{i}}^{r_{f}}[\int
\frac{2\sqrt{(\rho^2-\Delta')[(r^{2}+a^{2})^{2}-\Delta' a^{2}\sin
^{2}\theta ]}}{\Delta' \sqrt{\rho^2}}(dM'-\Omega'_HdJ')]dr.
\label{ims3}
\end{equation}
where
\begin{equation}\label{del}
\Delta^{^{\prime}}=r^{2}+a^{2}-2M'r
\end{equation}%

\begin{equation}\label{ri}
r_{i}=r_{+}=M+\sqrt{M^{2}-a^{2}},
\end{equation}%
\begin{equation}\label{rf}
r_{f}=M-\omega+\sqrt{(M-\omega)^{2}-a^{2}}.
\end{equation}
We see that $r=(M-\omega')+\sqrt{(M-\omega')^{2}-a^{2}}$ is a pole.
The integral can be evaluated by deforming the contour around the
pole, so as to ensure that positive energy solution decay in time.
Doing the $r$ integral first we obtain
\begin{equation}
{Im}S=-\frac{1}{2}\int_{(M)}^{(M-\omega)}
\frac{4\pi(M'^2+M'\sqrt{M'^{2}-a^{2}})}{\sqrt{M'^{2}-a^{2}}}
(dM'-\Omega'_HdJ')dr. \label{ims4}
\end{equation}
Finishing the integration we get
\begin{equation}
Im
S=\pi[M^2-(M-\omega)^2+M\sqrt{M^2-a^2}-(M-\omega)\sqrt{(M-\omega)^2-a^2}]=-\frac{1}{2}\Delta
S_{BH}.\label{ims5}
\end{equation}

The tunnelling rate is therefore
\begin{equation}
\Gamma\sim\exp[-2Im S]=e^{\Delta S_{BH}}. \label{spectrum}
\end{equation}
The Casimir energy Eq.(\ref{cas1}) now will be
 modified due to the semi-classical corrections as
 \begin{equation}\label{cascor}
E'_{C} = 2(3/2E' - T'S'_{BH}- \Omega'_{H} J').
\end{equation}
It is easily seen that \be \label{eqee}
2E'_{C}E'=\frac{V_{2}^{2}}{16\pi^{2}G^{2}r_{f}^{2}}(r_{f}^{2}+a^2)^{2}
\ee We substitute the above expression in the Cardy-Verlinde formula
in order that self-gravitational corrections to first order in
$\omega$ to be considered: \be
\label{cvcor}S'_{CFT}=S_{CFT}[1-\frac{2\omega
r_{i}^{2}}{(r_{i}^{2}+a^2)\sqrt{M^2-a^2}
}]=S_{CFT}[1-\frac{\omega(2M^2-a^2+2M\sqrt{M^2-a^2})}{(M^2+M\sqrt{M^2-a^2})\sqrt{M^2-a^2}}]
\ee For the Kerr black holes, the dual CFT lives on a flat space,
and thus the energy has no subextensive part. In the two-dimensional
conformal field theory, when the conformal weight of the ground
state is zero, we have \be S_{CFT}=2\pi \sqrt{\frac{cL_0}{6}},
\label{car2} \ee where $L_0=ER$ and
$\frac{c}{6}=\frac{2S_c}{\pi}=2E_cR$, here $S_c$ is the Casimir
entropy. Then, we can taking into account the self-gravitational
corrections of the Cardy-Verlinde entropy formula by just redefining
the Virasoro operator and the central charge as following. Using
Eqs.(13, 30, 31, 36) one can obtain

\be \label{lo} L'_{0}=E'R'=ER-\frac{\omega V_2  r_{i}^{2}}{4\pi G
\sqrt{M^2-a^2}}=L_0-\frac{\omega V_2
(2M^2-a^2+2M\sqrt{M^2-a^2})}{4\pi G \sqrt{M^2-a^2}} \ee \be
c'=12E'_{c}R'=12E_c R-\frac{6\omega V_2  r_{i}^{2}}{\pi G
\sqrt{M^2-a^2}}=c-\frac{6\omega V_2 (2M^2-a^2+2M\sqrt{M^2-a^2})}{\pi
G \sqrt{M^2-a^2}} \ee In \cite{ca} Carlip have computed the
logarithmic corrections to the Cardy formula, according to his
calculations, logarithmic corrections to the density of states is as
\be \label{ro} \rho(\Delta)\approx (\frac{c}{96
\Delta^{3}})^{1/4}\exp (2\pi \sqrt{\frac{c\Delta}{6}}) \ee where
$\Delta=L_0$ the exponential term in (\ref{ro}) gives the standard
Cardy formula, but we have now found the lading correction, which is
logarithmic. In the other hand as we saw the effect of the
self-gravitational corrections to the Cardy- Verlinde formula appear
as the redefinition of the $c$ and $L_0$ only. As Carlip have
discussed in \cite{ca}, the central charge $c$ appearing in
(\ref{ro}) is the full central charge of the conformal field theory.
In general, $c$ will consist of a classical term which appear in the
Poisson bracets of the Virasoro algebra generators, plus a
correction due to the quantum (here self-gravitational corrections)
effects, that can change the exponent in (\ref{ro}) from its
classical value. Moreover we saw that $L_0$ take a similar
correction as eq.(\ref{lo}) then the similar discussion about $L_0$
is correct. Therefore the first order corrections to the $L_0$ and
$c$ are given by \be \Delta L_0=L'_{0}-L_0=-\frac{\omega V_2
(2M^2-a^2+2M\sqrt{M^2-a^2})}{4\pi G \sqrt{M^2-a^2}}  \ee \be\Delta
c=c'-c=-\frac{6\omega V_2 (2M^2-a^2+2M\sqrt{M^2-a^2})}{\pi G
\sqrt{M^2-a^2}} \ee Thus, this redefinition can be considered as a
renormalization of the quantities entering in the Cardy formula.
 \section{Conclusions}
Black hole thermodynamic quantities depend on the event horizon
radius as Eqs.(\ref{tem}-\ref{angv}). The event horizon radius
undergoes corrections from the self-gravitational effect as
Eq.(\ref{rf}). Then we have obtained the corrections to the entropy
of a dual conformal field theory live on flat space as
Eqs.(\ref{cvcor}). Then we have considered this point that the
generalized Cardy-Verlinde formula is the outcome of a striking
resemblance between the thermodynamics of CFTs with asymptotically
flat dual's and CFTs in two dimensions. After that we have obtained
the corrections to the quantities entering the Cardy-Verlinde
formula:Virasoro operator and the central charge.

\end{document}